\documentclass[twocolumn,amsmath,amssymb,prd,preprintnumbers,letterpaper]{revtex4}

\usepackage{graphicx}
\usepackage{latexsym}
\usepackage{bm}

\def\al{\alpha}

\def\ve{\varepsilon}

\def\et{\eta}

\def\rh{\rho}

\def\si{\sigma}

\def\om{\omega}

\def\Th{\Theta}
\def\La{\Lambda}

\def\cl{{\cal L}}

\def\fr#1#2{{{#1} \over {#2}}}
\def\half{{\textstyle{1\over 2}}}
\def\quar{{\textstyle{1\over 4}}}

\def\lsim{\mathrel{\rlap{\lower3pt\hbox{$\sim$}}
    \raise2pt\hbox{$<$}}}
\def\gsim{\mathrel{\rlap{\lower3pt\hbox{$\sim$}}
    \raise2pt\hbox{$>$}}}
\def\sqr#1#2{{\vcenter{\vbox{\hrule height.#2pt
         \hbox{\vrule width.#2pt height#1pt \kern#1pt
         \vrule width.#2pt}
         \hrule height.#2pt}}}}

\def\prt{\partial}
\def\lrpartial{\raise 1pt\hbox{$\stackrel\leftrightarrow\partial$}}

\def\etal{{\it et al.}}

\newcommand{\bit}{\begin{itemize}}
\newcommand{\eit}{\end{itemize}}

\newcommand{\beq}[1]{\begin{equation}\label{#1}}
\newcommand{\eeq}{\end{equation}}
\newcommand{\bea}[1]{\begin{eqnarray}\label{#1}}
\newcommand{\eea}{\end{eqnarray}}
\newcommand{\ba}{\begin{array}}
\newcommand{\ea}{\end{array}}

\newcommand{\rf}[1]{(\ref{#1})}

\def\fr#1#2{{{#1} \over {#2}}}
\def\etal{{\it et al.}}

\begin{document}

\preprint{MIT-CTP-3797}

\title{Spacetime symmetries of the Lorentz-violating Maxwell--Chern--Simons model}

\author{A.J.\ Hariton}

\email[Electronic mail: ]{hariton@lns.mit.edu}
\affiliation{Center for Theoretical Physics,
Massachusetts Institute of Technology,
Cambridge, MA 02139}

\author{Ralf Lehnert}

\email[Electronic mail: ]{rlehnert@lns.mit.edu}
\affiliation{Center for Theoretical Physics,
Massachusetts Institute of Technology,
Cambridge, MA 02139}

\date{December 15, 2006}

\begin{abstract} 
The spacetime symmetries of 
classical electrodynamics 
supplemented with a Chern--Simons term 
that contains a constant nondynamical 4-vector 
are investigated. 
In addition to translation invariance 
and the expected three remaining Lorentz symmetries 
characterized by the little group of the external vector, 
the model possesses an additional spacetime symmetry 
if the nondynamical vector is lightlike. 
The conserved current associated 
with this invariance is determined, 
and the symmetry structure
arising from this invariance 
and the usual little group ISO(2)
is identified as SIM(2).
\end{abstract}


\maketitle

\section{Introduction}
\label{introduction}

Lorentz and CPT violation 
has recently received substantial attention 
as a potential signature for underlying physics, 
possibly arising from the Planck scale \cite{cpt04}. 
The prototype of a Lorentz- and CPT-breaking field theory 
consists of a three-dimensional Chern--Simons term 
embedded in Maxwell's four-dimensional classical electrodynamics \cite{CFJ}. 
For example, 
such a term 
is part of the Standard-Model Extension (SME), 
the general field-theory framework for Lorentz and CPT tests, 
which contains the Standard Model of particle physics 
and general relativity as limiting cases \cite{sme}. 
Numerous experimental and theoretical analyses 
of Lorentz and CPT breakdown have been performed 
within the SME. 

Astrophysical spectropolarimetry constrains
the Lorentz and CPT violation 
described by the Maxwell--Chern--Simons model 
to an extraordinary degree \cite{CFJ}, 
so that it can be set to zero for all practical purposes. 
Nevertheless, 
this model continues to enjoy a unique popularity for theoretical investigations, 
for it is simple, long-established, and mathematically interesting. 
Examples of recent studies in the context of the Maxwell--Chern--Simons model include ones involving 
radiative corrections \cite{radiative}, 
nontrivial spacetime topology \cite{klink}, 
causality \cite{caus}, 
energy positivity \cite{pos}, 
supersymmetry \cite{susy}, 
vacuum \v{C}erenkov radiation \cite{cer}, 
and the cosmic microwave background \cite{cmb}. 
More recently, 
the idea behind the construction of the Chern--Simons term in electrodynamics 
has also been applied to obtain a similar Lorentz-violating extension of general relativity \cite{CSG}. 

The present note continues along these lines 
and employs the Maxwell--Chern--Simons model 
as a simple theoretical laboratory 
to study the number of violated spacetime symmetries 
in the presence of an external 4-vector. 
Our results confirm 
that typically the Lorentz group is broken down to the little group 
associated with the external 4-vector. 
However, in certain circumstances 
involving the special case of a lightlike external vector, 
a further continuous symmetry relative to the timelike and spacelike cases 
may exist. 
In such situations, 
an additional conserved current exists, 
which often simplifies practical applications 
and may yield insight into the structure of the model. 

This paper is organized as follows.
Section \ref{model} reviews the basics of the Maxwell--Chern--Simons model. 
In Sec.\ \ref{symmetries}, 
we perform a general killing-vector analysis 
to identify the remaining spacetime symmetries of this model, 
and establish 
that a lightlike Lorentz violation maintains 
one more symmetry relative to the timelike and spacelike cases. 
The associated conserved current 
is constructed in Sec.\ \ref{current}. 
Section \ref{symmetrygroup} determines the resulting larger symmetry group. 
A brief summary is contained Sec.\ \ref{sum}.

\section{Photons with a Chern--Simons term}
\label{model}

This section 
reviews various results concerning 
the Lorentz- and CPT-violating Chern--Simons extension 
of electrodynamics. 
In natural units 
$c\hspace{-1pt} =\hspace{-1pt} \hbar\hspace{-1pt} =\hspace{-1pt} 1$,
the model Lagrangian
in the presence of external sources $j^{\mu}$
is given by 
\beq{lagr}
\cl_{\rm MCS} = 
-\fr{1}{4} F^2
+k_{\mu}A_{\nu}\tilde{F}^{\mu\nu}
-A\!\cdot\!j\;. 
\eeq
Here, 
$F_{\mu\nu}=\prt_{\mu}A_{\nu}-\prt_{\nu}A_{\mu}$ 
and its dual $\tilde{F}^{\mu\nu}=\half\ve^{\mu\nu\rh\si}F_{\rh\si}$ 
are defined as usual.
The nondynamical fixed $k^{\mu}$ 
determines a preferred direction in spacetime 
violating Lorentz as well as CPT symmetry. 
Although this Lagrangian is gauge dependent, 
the associated action integral is invariant 
if the source $j^{\mu}$ is conserved. 

The Lagrangian \rf{lagr} leads to the following
equations of motion for the potentials $A^{\mu}=(A^0,\vec{A})$: 
\beq{oddeom}
\left(\Box \et^{\mu\nu}-\prt^{\mu} \prt^{\nu}
-2\ve^{\mu\nu\rh\si}k_{\rh}\prt_{\si}\right)A_{\nu}
=j^{\mu} .
\eeq
As in conventional electrodynamics, 
current conservation $\prt_{\mu}j^{\mu}=0$ 
follows as a consistency requirement. 
The resulting modified Maxwell equations 
\beq{maxwell}
\partial_{\mu}F^{\mu\nu}+2k_{\mu}\tilde{F}^{\mu\nu}=j^{\nu} 
\eeq
are gauge invariant, 
as expected. 
For completeness, 
we also display the modified Coulomb 
and Amp\`ere laws contained in Eq.\ \rf{maxwell}: 
\bea{oddmax}
\vec{\nabla}\!\cdot\!\vec{E}-2\,\vec{k}\!\hspace{0.8pt}\cdot\!\vec{B} & = & \rh ,\nonumber\\
-\dot{\hspace{-1pt}\vec{E}}+\vec{\nabla}\!\times\!\vec{B}
-2\,k_0\vec{B}+2\,\vec{k}\!\times\!\vec{E} & = & 
\hspace{1.5pt}\vec{\hspace{-1.5pt}\textit{\j}}\hspace{1pt} .
\eea
The homogeneous Maxwell equations 
remain unaltered 
because the field--potential relationship is conventional. 

The Lagrangian \rf{lagr} is invariant under spacetime translations, 
and therefore a conserved energy--momentum tensor $\Th^{\mu\nu}$ can be constructed. 
Starting from the canonical expression 
and adding judiciously chosen 
superpotential terms $\partial_\alpha X^{[\alpha\mu]\nu}$, 
the relatively compact form 
\beq{emtensor} 
\Th^{\mu\nu}= 
\quar\,\et^{\mu\nu}F^2
-F^{\mu\al}F^{\nu}{}_{\!\al}
-k^{\nu}\tilde{F}^{\mu\al}A_{\al} 
\eeq
can be derived. 
Here, 
$\et^{\mu\nu}$ denotes the usual metric with signature $-2$. 
In the absence of sources $j^{\mu}=0$, 
this tensor is conserved in its $\mu$ index: $\partial_{\mu}\Theta^{\mu\nu}=0$. 
Note that the usual Belinfante symmetrization procedure 
is inapplicable 
because Lorentz symmetry is violated. 
Note also 
that $\Th^{\mu\nu}$ is gauge dependent. 
But the additional term $-k^{\nu}\tilde{F}^{\mu\al}\prt_\alpha\La=-\prt_\alpha k^{\nu}\tilde{F}^{\mu\al}\La$ 
generated by a gauge transformation
$A_{\alpha}\rightarrow A_{\alpha}+\prt_{\alpha}\La$ 
is a superpotential, 
which leaves unaffected the conserved 4-momentum $P^{\nu}\equiv\int d^3x\:\Th^{0\nu}$. 

The ansatz $A^{\mu}(x)=\epsilon^{\mu}(\lambda)\exp (-i\lambda\! \cdot\! x)$, 
where $\lambda^{\mu}\equiv(\omega,\vec{\lambda})$, 
together with the equations of motion \rf{oddeom} 
yields the plane-wave dispersion relation:
\beq{odddisp}
\lambda^4+4\,\lambda^2k^2-4\,(\lambda\!\hspace{0.8pt}\cdot\! k)^2=0 .
\eeq
This equation determines the wave frequency $\om$
for a given wave 3-vector $\vec{\lambda}$.
For a timelike $k^{\mu}$, 
the magnitude of the group velocity 
determined by the dispersion relation \rf{odddisp} 
can exceed the light speed $c$. 
Indeed, 
previous analyses have established theoretical difficulties 
associated with instabilities and causality violations 
for $k^2>0$. 
These issues are absent for $k^2\leq 0$. 
In what follows, 
we focus primarily on the case of a lightlike $k^{\mu}$ in the absence of sources $j^{\mu}=0$.

\section{Remaining spacetime symmetries} 
\label{symmetries} 

Besides the usual ten Poincar\'e invariances 
associated with four translations, 
three rotations, 
and three boosts, 
the conventional free Maxwell field possesses five additional spacetime symmetries 
arising from one dilatation 
and four special conformal transformations. 
The inclusion of the Chern--Simons term ${\cal L}_{\rm CS}\equiv k^\alpha\, A^\beta\, \tilde{F}_{\alpha\beta}$ 
maintains translation invariance, 
since $k^{\mu}$ is assumed to be constant. 
However, 
one expects 
that dilatation and conformal symmetry are lost 
because $k^{\mu}$ has mass dimensions 
setting a definite scale. 
One further expects 
that the Lorentz group is broken down to the appropriate little group 
associated with $k^{\mu}$. 
For timelike, spacelike, and lightlike $k^{\mu}$, 
the little groups are SO(3), SO(2,1), and ISO(2), 
respectively. 
Each of these groups is three dimensional, 
so that at least three of the original six Lorentz symmetries are maintained. 
This section shows 
that for a lightlike $k^{\mu}$ 
one additional spacetime symmetry, 
which is a combination of a dilatation and a boost, 
exists. 

We begin by recalling 
that a conformal Killing-vector field $f^{\mu}(x)$
associated with the four-dimensional Minkowski metric $\eta^{\mu\nu}$ 
satisfies 
\beq{KillingDef}
\partial^{\mu}f^{\nu}+\partial^{\nu}f^{\mu}=\half\,\eta^{\mu\nu}\,\partial_{\alpha}f^{\alpha}\;,
\eeq
and possesses the general solution 
\beq{GenKilling}
f^{\mu}=a^{\mu}+\omega^{\mu\nu}x_{\nu}+\rho\,x^{\mu}+2\,(x\!\cdot\!b)\,x^{\mu}-x^2\,b^{\mu}\;.
\eeq
Here, 
$a^{\mu}$, $\omega^{\mu\nu}=-\omega^{\nu\mu}$, $c$, and $b^{\mu}$ 
are free coefficients 
parametrizing translations, Lorentz transformations, dilatations, and conformal transformations, 
respectively. 
With this Killing-vector field we may construct the current
\beq{current}
J^{\mu}\equiv\Theta^{\mu\nu}f_\nu\;.
\eeq
Employing the defining Eq.\ \rf{KillingDef} 
and  energy--momentum conservation $\partial_\mu\Theta^{\mu\nu}$, 
one finds that $\partial_{\mu}J^{\mu}=\quar\, \Theta^{\mu}{}_{\mu}\,\partial_{\alpha}f^{\alpha}+\Theta^{\mu\nu}\,\partial_{[\mu}f_{\nu]}$. 
In the conventional Maxwell case, 
this divergence vanishes 
because $\Theta^{\mu\nu}$ is traceless and symmetric 
confirming the existence of 15 conserved quantities. 

In the present Lorentz-violating case, 
$\Theta^{\mu\nu}$ possesses a piece 
that fails to be traceless and symmetric: 
the term containing $k^{\nu}$ in Eq.\ \rf{emtensor}. 
The Chern--Simons extension therefore leads to 
\begin{widetext}
\beq{divJ} 
\partial_{\mu}J^{\mu}=
\big[\omega_{\mu\nu}k^{\nu}-\rho\,k_{\mu}
+2\,(k\!\cdot\!b)\,x_{\mu}-2\,(k\!\cdot\!x)\,b_{\mu}-2\,(b\!\cdot\!x)\,k_{\mu}\big] \tilde{F}^{\mu\alpha}A_{\alpha}\;.
\eeq 
\end{widetext}
This is conserved for arbitrary solutions $A_{\alpha}$, $\tilde{F}^{\mu\alpha}$ 
and for arbitrary gauges, 
if the expression in the brackets vanishes identically. 
This expression can be viewed as a linear function in the free variable $x$, 
so that two simultaneous conditions emerge: 
\beq{cond1} 
(k\!\cdot\!b)\,\eta^{\mu}_{\nu}-b^{\mu}k_{\nu}-k^{\mu}b_{\nu}=0
\eeq
from the coefficient in front of $x$ and 
\beq{cond2}
\omega_{\mu\nu}k^{\nu}-c\,k_{\mu}=0
\eeq
from the $x$-independent term. 
These conditions determine the remaining symmetries. 

Condition \rf{cond1} implies 
that $b^{\mu}=0$, 
which can be established as follows. 
The trace of Eq.\ \rf{cond1} 
gives $k\!\cdot\!b=0$. 
Using this fact 
and contracting Eq.\ \rf{cond1} with $b^{\nu}$ 
shows that $b^2=0$. 
Similarly, 
contraction with $k^{\nu}$ yields $b^{\mu}k^2=0$. 
Suppose $b^{\mu}\neq0$, 
so that $k^2=0$. 
Then, 
the requirements $k\!\cdot\!b=k^2=b^2=0$ 
imply that $b^{\mu}$ and $k^{\mu}$ are (anti)parallel. 
But then, 
the condition \rf{cond1} 
could be cast into the form $k^{\mu}k_{\nu}=0$, 
which is inconsistent with the assumption of a nontrivial Chern--Simons term. 
It follows that $b^{\mu}$ must indeed vanish, 
so that conformal symmetry is broken 
in our Maxwell--Chern--Simons model, 
as expected. 

Condition \rf{cond2} 
can be satisfied for $\rho=0$ 
and $\omega^{\mu\nu}k_{\nu}=0$. 
This means 
that $\omega^{\mu\nu}$ must lie in the parameter space of the little group 
associated with $k^{\mu}$. 
For example, 
if $k^{\mu}$ is timelike, 
we may select an inertial coordinate system in which $k^{\mu}=(k^0,\vec{0})$. 
In this frame, 
we need $\omega^{0j}=-\omega^{j0}=0$ to ensure $\omega^{\mu\nu}k_{\nu}=0$, 
i.e., boosts are no longer a symmetry. 
Here, 
spatial components are denoted by lower-case Latin indices $j,k,l,$ etc.
However, 
arbitrary spatial rotations 
parametrized by $\omega^{jk}=\epsilon^{jkl}\theta^{l}$, 
which correspond to the little group SO(3), 
remain compatible with $\omega^{\mu\nu}k_{\nu}=0$. 
A similar reasoning applies 
to the cases of a spacelike and lightlike $k^{\mu}$ 
with their respective little groups SO(2,1) and ISO(2). 
This $\rho=0$ solution to the condition \rf{cond2} 
is expected and unsurprising. 

We may also ask 
whether condition \rf{cond2} 
can be satisfied for the case $c\neq 0$. 
Contraction of Eq.\ \rf{cond2} 
with $k^{\mu}$ shows 
that $k^2=0$ is a necessary requirement 
in this case. 
Separating the temporal and spatial components of condition \rf{cond2} 
yields
\beq{temp2}
\vec{\beta}\!\cdot\hat{k}=c
\eeq
and 
\beq{spat2}
\vec{\beta}+\vec{\theta}\times\hat{k}=c\,\hat{k}
\eeq
Here, 
we have defined $k^{\mu}=k(1,\hat{k})$, 
where $\hat{k}$ is a unit 3-vector. 
In addition, 
we have denoted the rapidity $\vec{\beta}$ of a boost by $\omega_{0j}=\beta^{j}$ 
and the angle $\vec{\theta}$ of a rotation by $\omega_{jk}=\omega^{jk}=\epsilon^{jkl}\theta^l$. 
Equation \rf{spat2} 
may be further decomposed
into its components
parallel and perpendicular to $\hat{k}$: 
\begin{subequations}
\begin{eqnarray}
\vec{\beta}_\parallel & = & c\,\hat{k}\;,\label{spatdecomp1}\\ 
\vec{\beta}_\perp & = & \hat{k}\times\vec{\theta}\;.\label{spatdecomp2}
\end{eqnarray}
\end{subequations}
Here, 
we have set $\vec{\beta}_\parallel\equiv\vec{\beta}\!\cdot\hat{k}$ 
and $\vec{\beta}_\perp\equiv\vec{\beta}-\vec{\beta}\!\cdot\hat{k}$. 
We remark in passing 
that Eq.\ \rf{spatdecomp1} 
is equivalent to Eq.\ \rf{temp2}, 
which means Eq.\ \rf{temp2} is also contained in Eq.\ \rf{spat2}. 
Equation \rf{spatdecomp2}, 
which does not contain $\rho$, 
is associated with the expected 
three remaining Lorentz symmetries 
described the little group ISO(2) of our lightlike $k^{\mu}$.
One of these corresponds to rotations $\vec{\theta}=\theta\,\hat{k}$ about $\hat{k}$. 
The other two 
correspond to the two possible independent boosts in the plane orthogonal to $\hat{k}$, 
which must be simultaneously performed together with the appropriate rotation about an axis 
perpendicular to both the boost direction and $\hat{k}$. 
In addition to this standard result for the little group of a lightlike 4-vector, 
there is one more symmetry in the present case: 
according to Eq.\ \rf{temp2}, or equivalently Eq.\ \rf{spatdecomp1}, 
a boost along the direction of $\hat{k}$ together with an appropriate dilatation 
also satisfies condition \rf{cond2}. 
In the next section, 
we discuss this additional invariance further.

\section{Additional spacetime symmetry for lightlike $\bm k^{\mu}$} 
\label{current} 

We begin by recalling 
that a dilatation,
which is also called a scale transformation, 
takes $x^{\mu}\to x_D^{\mu}=e^{+\rho}x^{\mu}$ 
and $A^{\mu}\to A_D^{\mu}=e^{-\rho}A^{\mu}$, 
where the size of the dilatation is determined 
by the parameter $\rho$. 
It is apparent 
that the unconventional Chern--Simons-type term in Lagrangian \rf{lagr} 
not only violates Lorentz symmetry, 
but it also breaks scale invariance 
because $k^{\mu}$ has mass dimensions. 
To see this explicitly, 
we decompose the Lagrangian \rf{lagr} 
according to ${\cal L}={\cal L}_{\rm M}+{\cal L}_{\rm CS}$. 
Here,  
\beq{convdef}
{\cal L}_{\rm M}=-\quar\, F^2 
\eeq
denotes the conventional Maxwell piece and  
\beq{CSdef}
{\cal L}_{\rm CS}=k^\alpha\, A^\beta\, \tilde{F}_{\alpha\beta} 
\eeq 
the Chern--Simons piece, 
as before.
We remind the reader 
that we consider the free case $j^{\mu}=0$ only. 
A dilatation takes 
\beq{dil}
{\cal L} \to {\cal L}_{\rm M}+e^{-\rho}\,{\cal L}_{\rm CS} \neq {\cal L}\;. 
\eeq
Note that $x$ becomes a dummy integration variable in the action, 
so the $x$ dependence of the fields in the above transformation can be suppressed. 
We see 
that the conventional piece and the Chern--Simons extension transform differently. 
Moreover, 
the difference between the original and the transformed Lagrangians 
fails to be a total derivative, 
which establishes the non-invariance of ${\cal L}$ under dilatations. 

We next consider Lorentz transformations, 
which can be implemented via $\Lambda^{\mu}{}_{\nu}(\vec{\theta},\vec{\beta})$. 
As before, 
$\vec{\theta}$ and $\vec{\beta}$ characterize rotations and boosts, respectively. 
Under such transformations, 
the Lagrangian \rf{lagr} changes according to 
${\cal L} \to {\cal L}_{\rm M}+ \Lambda^{\mu}{}_{\gamma}(-\vec{\theta},-\vec{\beta})\,k^{\gamma}\,A^{\nu}\,\tilde{F}_{\mu\nu}$
in the absence of sources.
We have again suppressed the dependence on the dummy integration variables $x$ 
for brevity. 
Motivated by the discussion in the previous section, 
we consider a boost along $\hat{k}$ with rapidity $\beta$. 
Such a transformation changes the magnitude of $k^{\mu}$ by a factor of $e^\beta$. 
We then have 
$\Lambda^{\mu}{}_{\gamma}(\vec{0} ,-\beta\hat{k})\,k^{\gamma}\,A^{\nu}\,\tilde{F}_{\mu\nu}=\exp(\beta)\, k^\mu\, A^\nu\, \tilde{F}_{\mu\nu}\neq{\cal L}_{\rm CS}$,
so that 
\beq{LT}
{\cal L} \to {\cal L}_{\rm M}+e^{\beta}\,{\cal L}_{\rm CS} \neq {\cal L}\;, 
\eeq
which establishes that symmetry under boosts along $\hat{k}$ is violated, 
as expected. 

Although each individual transformation \rf{dil} and \rf{LT} 
is no longer associated with a symmetry, 
the specific form of these transformations 
and the arguments in the previous section 
show 
that a dilatation {\it combined} with 
a suitable boost along the spatial direction of a lightlike $k^{\mu}$ 
remains a symmetry of the free part of Lagrangian \rf{lagr}. 
We can see this explicitly 
by examining the currents 
\beq{Dcurrent}
D^{\mu}\equiv\theta^{\mu\nu}x_{\nu}
\eeq 
and 
\beq{Lcurrent}
J^{\mu}_{\alpha\beta}\equiv \theta^{\mu}_{\alpha}x_{\beta}-\theta^{\mu}_{\beta}x_{\alpha}\;.
\eeq 
These quantities are defined 
to give the usual dilatation and Lorentz currents 
in the $k^{\mu}\to 0$ limit. 
To extract from Eq.\ \rf{Lcurrent} 
the current associated with a boost along $\hat{k}$, 
we split $k^{\mu}$ into its purely timelike and its purely spacelike part 
$k^{\mu}=k\,(k_T^{\mu}+k_S^{\mu})$, 
where $k_T^{\mu}=(1,\vec{0})$ and $k_S^{\mu}=(0,\hat{k})$. 
The projection onto the desired components 
is then given by $J^{\mu}_{\alpha\beta}\,k_S^{\alpha}\,k_T^{\beta}$. 
The divergences of these currents 
obey 
\beq{Ddiv}
\partial_{\mu}D^{\mu}=-{\cal L}_{\rm CS}
\eeq
and 
\beq{Ldiv}
\partial_{\mu}J^{\mu}_{\alpha\beta}\,k_S^{\alpha}\,k_T^{\beta}=+{\cal L}_{\rm CS}\;. 
\eeq 
It is again apparent 
that $D^{\mu}$ and $J^{\mu}_{\alpha\beta}\,k_S^{\alpha}\,k_T^{\beta}$ 
fail to be conserved individually, 
but their sum $Q^{\mu}\equiv D^{\mu}+J^{\mu}_{\alpha\beta}\,k_S^{\alpha}\,k_T^{\beta}$ 
determines, in fact, a conserved current. 
An explicit expression for $Q^{\mu}$ 
can be obtained via their definitions \rf{Dcurrent} and \rf{Lcurrent} 
as well as Eq.\ \rf{emtensor}: 
\beq{explexpr}
Q^{\mu}=\big[\quar\, \eta^{\mu}_{\nu}F^2+F^{\mu\alpha}F_{\alpha\nu}\big]\big[x_{\nu}+(k_T\!\cdot\! x)\,k_S^{\nu}-(k_S\!\cdot\! x)\,k_T^{\nu}\big]\;.
\eeq
This expression puts into evidence 
the manifest gauge invariance of $Q^{\mu}$.

\section{Associated spacetime-symmetry group}
\label{symmetrygroup} 

In the previous sections, 
we have found an additional symmetry for the Maxwell--Chern--Simons model 
with lightlike Lorentz violation.
It is now natural to ask 
what the full spacetime-symmetry group 
of this model is. 
This question is the subject of the present section.

In the case of a lightlike $k^{\mu}$, 
the little group 
(i.e., the unbroken subgroup of the Lorentz group) 
is isomorphic to the three-dimensional Euclidean group ISO(2), 
which consists of rotations and translations in two dimensions. 
Since it is always possible to find an inertial coordinate system 
in which the vector $k^{\mu}$ points along the $z$-axis, 
we may assume this choice without loss of generality.
The group ISO(2)is then generated by 
the Lie algebra spanned by the following generators:
$A = J_2+K_1$, $B=-J_1+K_2$, and $J_3$. 
Here, 
$J_1$, $J_2$, $J_3$ denote the generators of rotations 
about the $x$, $y$, and $z$ axes respectively, 
and $K_1$, $K_2$, $K_3$ generate boosts along the $x$, $y$, and $z$ axes. 

The additional symmetry found in the previous sections corresponds to 
a combination of a dilation with a boost in the spatial direction of $k^{\mu}$. 
This transformation can be generated by an element of the form $C=K_3+D$, 
where $D$ is the usual uniform dilatation in the independent and dependent variables. 
Because of its uniformity, 
the dilation $D$ commutes with all of the rotation and boost operators listed above. 
Besides translations, 
the spacetime-symmetry group for the Maxwell--Chern--Simons model with lightlike symmetry breaking 
is thus generated by 
\beq{sim2gen}
A = J_2+K_1\;,\quad B=-J_1+K_2\;,\quad J_3\;,\quad C = K_3+D\;, 
\eeq
where the first three operators 
are the generators of ISO(2). 

For further study of the structure of this symmetry, 
we determine the commutation relations between the generators \rf{sim2gen}. 
To this end, 
we recall the commutation relations between elements of the usual Lorentz algebra: 
\begin{subequations}
\begin{eqnarray}
\left[J_j,J_k\right] & = & i\varepsilon^{jkl}J_l\;,\label{lrza1}\\
\left[J_j,K_k\right] & = & i\varepsilon^{jkl}K_l\;,\label{lrza2}\\
\left[K_j,K_k\right] & = & -i\varepsilon^{jkl}J_l\;.\label{lrza3}
\end{eqnarray}
\end{subequations}
These equations
and the fact 
that $D$ commutes with all $J_j$ and $K_j$  
determine the commutation relations of the vector fields \rf{sim2gen}, 
which we have summarized in Table \ref{table}.
\begin{table}[htbp]
\label{table}
  \begin{center}
\caption{\label{table}Commutation table for the Lie algebra spanned by the
  vector fields (\ref{sim2gen}).}
\vspace{5mm}
\begin{tabular}{|c||c|c|c|c|}\hline
 & $\mathbf{A}$ & $\mathbf{B}$ & $\mathbf{J_3}$ & $\mathbf{C}$ \\[0.5ex]\hline\hline
$\mathbf{A}$ & $0$ & $0$ & $-iB$ & $iA$ \\\hline
$\mathbf{B}$ & $0$ & $0$ & $iA$ & $iB$ \\\hline
$\mathbf{J_3}$ & $iB$ & $-iA$ & $0$ & $0$ \\\hline
$\mathbf{C}$ & $-iA$ & $-iB$ & $0$ & $0$ \\\hline
\end{tabular}
  \end{center}
\end{table}

Inspection of the Commutation Table \ref{table} reveals
that the algebra closes. 
Moreover, 
these commutation relations can be identified with those of the abstract Lie algebra sim(2) 
of the four-dimensional similitude group of $\mathbb{R}^2$, 
which is explicitly given by $\textrm{SIM(2)} = \mathbb{R}^+\times\mathbb{R}^2\times \textrm{SO(2)}$ \cite{sim2}. 
In the conventional notation \cite{not}, 
the algebra sim(2) is generated by the vector fields $T_a$, $T_{b_1}$, $T_{b_2}$, $T_{\theta}$, 
and its commutation relations are given by
\beq{simcommut}
[T_a,T_{b_k}] = -iT_{b_k}\;,\quad [T_{\theta},T_{b_k}]=-i\varepsilon^{3kl}T_{b_l}\;.
\eeq
Comparison with the Commutation Table \ref{table} establishes 
that we may identify $T_a$ with $C$, $T_{b_1}$ with $B$, $T_{b_2}$ with $A$, and $T_{\theta}$ with $J_3$. 

We finally remark 
that the similitude group SIM(2) 
has recently been employed as 
the starting point for a particular approach to Lorentz violation \cite{cohen}. 
From the perspective of this approach, 
the present Maxwell--Chern--Simons model 
represents a specific realization of a SIM(2)-invariant theory. 
As opposed to the fermion example considered in the SIM(2) approach, 
our SIM(2)-invariant model does not involve nonlocal operators. 
We note 
that the SIM(2) approach is compatible 
with a subset of the usual supersymmetries \cite{freedman}.

\section{Summary}
\label{sum}

This work has investigated the spacetime symmetries 
in the free Lorentz- and CPT-violating Maxwell--Chern--Simons model. 
The number of these symmetries depends on the spacetime character 
of the background vector producing the symmetry breaking: 
for a lightlike vector one more invariance 
relative to the timelike and spacelike cases 
exists.  
This additional symmetry 
results from a specific combination of a boost and a dilatation. 
We have determined the associated conserved current, 
which is given in Eq.\ \rf{explexpr}. 
In Sec.\ \ref{symmetrygroup}, 
we have demonstrated 
that the usual little group ISO(2) 
associated with a lightlike vector 
is enlarged to the similitude group SIM(2)  
by this additional symmetry. 
We expect similar results to hold 
in other scale invariant models 
that are supplemented only by Lorentz violation
with a single lightlike direction.

\acknowledgments
The authors thank Roman Jackiw for discussion. 
This work is supported 
by the U.S.\ Department of Energy 
under cooperative research agreement No.\ DE-FG02-05ER41360. 
A.J.H. acknowledges support by the FQRNT of Canada 
under their postdoctoral research fellowship program. 
R.L.\ is supported by the European Commission 
under Grant No.\ MOIF-CT-2005-008687.

\end{document}